\documentclass[aps,prl,twocolumn,showpacs,superscriptaddress,groupedaddress]{revtex4}  
\usepackage{graphicx}  
\usepackage{dcolumn}   
\usepackage{bm}        
\usepackage{amssymb}   

\newcommand{\etal}{{\it et~al.}}
\newcommand{\eg}{{\it e.g.}}

\begin{document}

\preprint{}

\title{Full Electroresistance Modulation in a Mixed-Phase Metallic Alloy}

\author{Z. Q. Liu}
\altaffiliation[Email: ]{liuzhiqi@berkeley.edu}
\affiliation{Center for Nanophase Materials Sciences, Oak Ridge National Laboratory, Oak Ridge, TN 37831, USA}
\affiliation{Department of Materials Science and Engineering, University of California, Berkeley, CA 94720, USA}

\author{L. Li}
\affiliation{Materials Science and Technology Division, Oak Ridge National Laboratory, Oak Ridge, TN 37831, USA}

\author{Z. Gai}
\affiliation{Center for Nanophase Materials Sciences, Oak Ridge National Laboratory, Oak Ridge, TN 37831, USA}

\author{J. D. Clarkson}
\author{S. L. Hsu}
\affiliation{Department of Materials Science and Engineering, University of California, Berkeley, CA 94720, USA}

\author{A. T. Wong}
\affiliation{Materials Science and Technology Division, Oak Ridge National Laboratory, Oak Ridge, TN 37831, USA}
\affiliation{Materials Science and Engineering, University of Tennessee, Knoxville, TN 37996, USA}

\author{L. S. Fan}
\affiliation{Materials Science and Technology Division, Oak Ridge National Laboratory, Oak Ridge, TN 37831, USA}

\author{M.-W. Lin}
\author{C. M. Rouleau}
\affiliation{Center for Nanophase Materials Sciences, Oak Ridge National Laboratory, Oak Ridge, TN 37831, USA}

\author{T. Z. Ward}
\author{H. N. Lee}
\affiliation{Materials Science and Technology Division, Oak Ridge National Laboratory, Oak Ridge, TN 37831, USA}

\author{A. S. Sefat}
\affiliation{Materials Science and Technology Division, Oak Ridge National Laboratory, Oak Ridge, TN 37831, USA}

\author{H. M. Christen}
\altaffiliation[Email: ]{christenhm@ornl.gov}
\affiliation{Center for Nanophase Materials Sciences, Oak Ridge National Laboratory, Oak Ridge, TN 37831, USA}

\author{R. Ramesh}
\altaffiliation[Email: ]{rramesh@berkeley.edu}
\affiliation{Department of Materials Science and Engineering, University of California, Berkeley, CA 94720, USA}
\affiliation{Department of Physics, University of California, Berkeley, CA 94720, USA}
\affiliation{Materials Sciences Division, Lawrence Berkeley National Laboratory, Berkeley, CA 94720, USA}
\affiliation{Oak Ridge National Laboratory, Oak Ridge, TN 37831, USA}

\date{\today}

\begin{abstract}
We report a giant, $\sim$22\%, electroresistance modulation for a metallic alloy above room temperature. It is achieved by a small electric field of 2 kV/cm via piezoelectric strain-mediated magnetoelectric coupling and the resulting magnetic phase transition in epitaxial FeRh/BaTiO$_{3}$ heterostructures. This work presents a detailed experimental evidence for an isothermal magnetic phase transition driven by tetragonality modulation in FeRh thin films, which is in contrast to the large volume expansion in the conventional temperature-driven magnetic phase transition in FeRh. Moreover, all the experimental results in this work illustrate FeRh as a mixed-phase model system well similar to phase-separated colossal magnetoresistance systems with phase instability therein.

\end{abstract}



\maketitle

Phase instability near a phase transition (PT) has been an intriguing subject owing to coexisting phases interacting and competing with each other [1]. Around the PT, multiple phases are comparable in terms of the free energy and thus the phase transformation among them could be triggered by even subtle energetic excitations. Accordingly, it can result in various fascinating physical phenomena, such as enhancement of superconducting transition temperature (\emph{T}) [2], colossal magnetoresistance (CMR) [3-6], giant electroresistance and magnetoelectric coupling related to manganites [7-9], and large electric field (\emph{E})-induced strain around phase boundaries in a ferroelectric material [10].

FeRh, a metallic binary alloy, can exhibit a chemically-ordered $\emph{bcc}$-B2 (CsCl-type) $\alpha$-phase or a chemically-disordered $\emph{fcc}$ $\gamma$-phase at room temperature in bulk around the equiatomic concentration [11]. The $\emph{fcc}$ phase is nonmagnetic, while the \emph{bcc} phase undergoes a first-order magnetic phase transition (MPT) from $\emph{G}$-type antiferromagnetic (AFM) to ferromagnetic (FM) order upon heating above a transition \emph{T} between 350$\sim$370 K [12-14]. The transition is accompanied by a volume increase of $\sim$1\% with the \emph{bcc} structure retained [15]. Meanwhile, a resistivity drop [14] occurs as the FM phase has a lower resistivity, which may be closely related to electronic structure changes [16] and spin disorder present in the system [17]. The origin of the \emph{T}-driven MPT is still under debate. Kittel originally proposed a thermodynamic model of exchange-inversion induced by lattice expansion for \emph{T}-driven MPT systems [18]. In addition, important roles of electronic entropy [19], instability of the rhodium moment [20], spin wave excitations [21], and magnetic excitations [22-24] have all been emphasized for FeRh.

Despite of the mechanism being controversial, the research on FeRh has been stimulated by its potential applications in various information storage media based on the \emph{T}-driven MPT. Thiele $\etal$  demonstrated that FeRh can be utilized as thermally assisted magnetic recording media in 2003 [25]. Subsequently in 2004, both Ju $\etal$ [26] and Thiele $\etal$ [27] reported that a femtosecond laser pulse drives FeRh from the AFM phase to the FM phase as a consequence of laser heating. Furthermore, the use of FeRh in a room-temperature AFM memory resistor has been recently demonstrated [28].

In addition to the conventional \emph{T}-driven MPT, emerging epitaxial integration of FeRh thin films with functional ferroelectric oxides has enabled an isothermal \emph{E}-controlled MPT [29-31] via piezoelectric strain. Recently studies [29,31] on FeRh/BaTiO$_{3}$ (BTO) heteostructures have achieved the MPT in FeRh thin films using \emph{E}-generated piezoelectric strain in BTO single crystals and highly efficient strain mediation at epitaxial FeRh/BTO interfaces. Direct magnetic measurements with \emph{E}-fields reveal a giant magnetoelectric coupling coefficient of $\alpha$ = $\mu$$_{0}$$\Delta$$\emph{M}$/$\Delta$$\emph{E}$ $\thicksim$ 10$^{-5}$ s m$^{-1}$ [29], which is over four orders of magnitude larger than in single-phase multiferroic materials (10$^{-12}$ s m$^{-1}$ for Cr$_{2}$O$_{3}$ [32], 10$^{-10}$ s m$^{-1}$ for TbMnO$_{3}$ [33], and 10$^{-9}$ s m$^{-1}$ for Ni$_{3}$B$_{7}$O$_{13}$I [34]), and one or two orders of magnitude greater than the reported values for multiferroic heterostructures (10$^{-7}$ s m$^{-1}$ for La$_{0.67}$Sr$_{0.33}$MnO$_{3}$/BTO [9], and 10$^{-6}$ s m$^{-1}$ for Co$_{40}$Fe$_{40}$B$_{20}$/Pb(Mg$_{1/3}$Nb$_{2/3}$)$_{0.7}$Ti$_{0.3}$O$_{3}$ [35]). It has been further shown that such isothermal strain-induced MPT works and is more reversible at the nanoscale [31]. Our previous work [30] instead has focused on electrical transport of FeRh thin films during \emph{E}-induced magnetic phase change in epitaxial FeRh/0.72PbMg$_{1/3}$Nb$_{2/3}$O$_{3}$-0.28PbTiO$_{3}$ (PMN-PT) heterostructures and achieved an $\sim$8\% resistivity modulation in FeRh, which is potential for nonvolatile resistor memory devices.

Although previous studies [29-31] have generally demonstrated the isothermal \emph{E}-controlled MPT in FeRh thin films, the decisive factor for driving the MPT is still an open question. As one may expect from Kittel's model [18] and the \emph{T}-driven MPT, the volume change should be responsible for the MPT in such strain-mediated magnetoelectric heterostructures. However, in the FeRh/BTO work [29,31], the out-of-plane lattice constant of FeRh was experimentally measured, but the in-plane lattice constant was estimated via the Poisson's ratio of bulk FeRh $\nu$ = 0.31 [29]. The Poisson's ratio could be largely different between free-standing bulk and epitaxial thin films owing to epitaxial strain and distinct mechanical properties. Therefore, the information on the actual volume change was absent. On the other hand, our previous work [30] examined the lattice change of FeRh with \emph{E}-fields only at room temperature, which may have not shed sufficient light on the \emph{E}-controlled MPT occurring at high temperatures. Moreover, the in-plane compressive strain obtained in FeRh/PMN-PT heterostructures was -0.075\%, smaller than in FeRh/BTO, leading to only partial magnetic phase switch in FeRh films as revealed by the magnetic force microscopy (MFM) measurements [30].

In this letter, we focus on electrical transport of epitaxial FeRh/BTO heterostructures. Together with MFM and structural measurements with $\emph{in-situ}$ \emph{E}-fields, the work is to investigate how an external \emph{E} changes the magnetic phase via strain in FeRh films. We achieve a resistivity modulation of $\sim$22\% for FeRh films with 2 kV/cm applied across the BTO substrate, corresponding to the full magnetic phase switch, which is predominantly induced by the modulation of tetragonality in FeRh films rather than volume variation as in the \emph{T}-driven MPT. Moreover, the 22\% resistivity modulation is giant, for single metals or metal alloys, which has benefited from the phase instability around the PT in FeRh.

In this work, FeRh thin films were grown on BTO single-crystal substrates (0.5 mm thick, purchased from MTI Corp., US) by a d.c. sputtering system with a base pressure of 2$\times$10$^{-8}$ Torr. A stoichiometric Fe$_{50}$Rh$_{50}$ sputtering target with a diameter of 5.08 cm and a thickness of 3 mm was used. Considering that titanates such as SrTiO$_{3}$ and BTO have high oxygen diffusion coefficients at high temperatures [36,37], the fabrication was kept at relatively low temperatures to avoid interfacial oxidation: growth at 375 $^\circ$C and annealing at 520 $^\circ$C for 1 h. The sputtering power and Ar pressure were optimized to be 50 W and 5 mTorr, respectively, giving rise to Fe$_{49.5}$Rh$_{50.5}$ thin films. The growth rate was determined to be 1.54 nm/min. To avoid cracking BTO single crystals across the structural PT at $\sim$130 $^\circ$C, the ramping rate was kept at a low value during both heating and cooling, $\emph{i.e.}$, 2 $^\circ$C/min.

Structural and magnetic characterizations of a highly epitaxial 35-nm-thick FeRh/BTO heterostructure are in the supplemental material (Figs. S1 \& S2). High resolution (001) and (101) $\theta$-2$\theta$ scans of the FeRh peaks yield \emph{a} = 2.9725 {\AA} and \emph{c} = 3.0050 {\AA} for the FeRh film, corresponding to the tetragonality \emph{c}/\emph{a} = 1.0110. Compared with the \emph{bcc} structure in bulk (\emph{a}$\sim$2.99 {\AA} [14]), the thin film is compressively strained to be tetragonal due to lattice mismatch. The magnetic transition sets in at $\sim$350 K during warming [Fig. S1(d)] in the temperature-dependent magnetization \emph{M-T} curve. The thermal hysteresis during warming and cooling demonstrates the nature of a first-order MPT. The epitaxial FeRh film exhibits a magnetization of $\sim$50 emu/cc even below 300 K, which is larger than that of the AFM phase in bulk ($<$ 10 emu/cc). This is likely induced by interfacial ferromagnetism existing in AFM FeRh epilayers as observed in FeRh/MgO heterostructures [38]. A weak kink at $\sim$288 K is present in both the warming and cooling \emph{M-T} curves, signature of the BTO tetragonal to orthorhombic structural PT.

\begin{figure}
\includegraphics[width=3.4in]{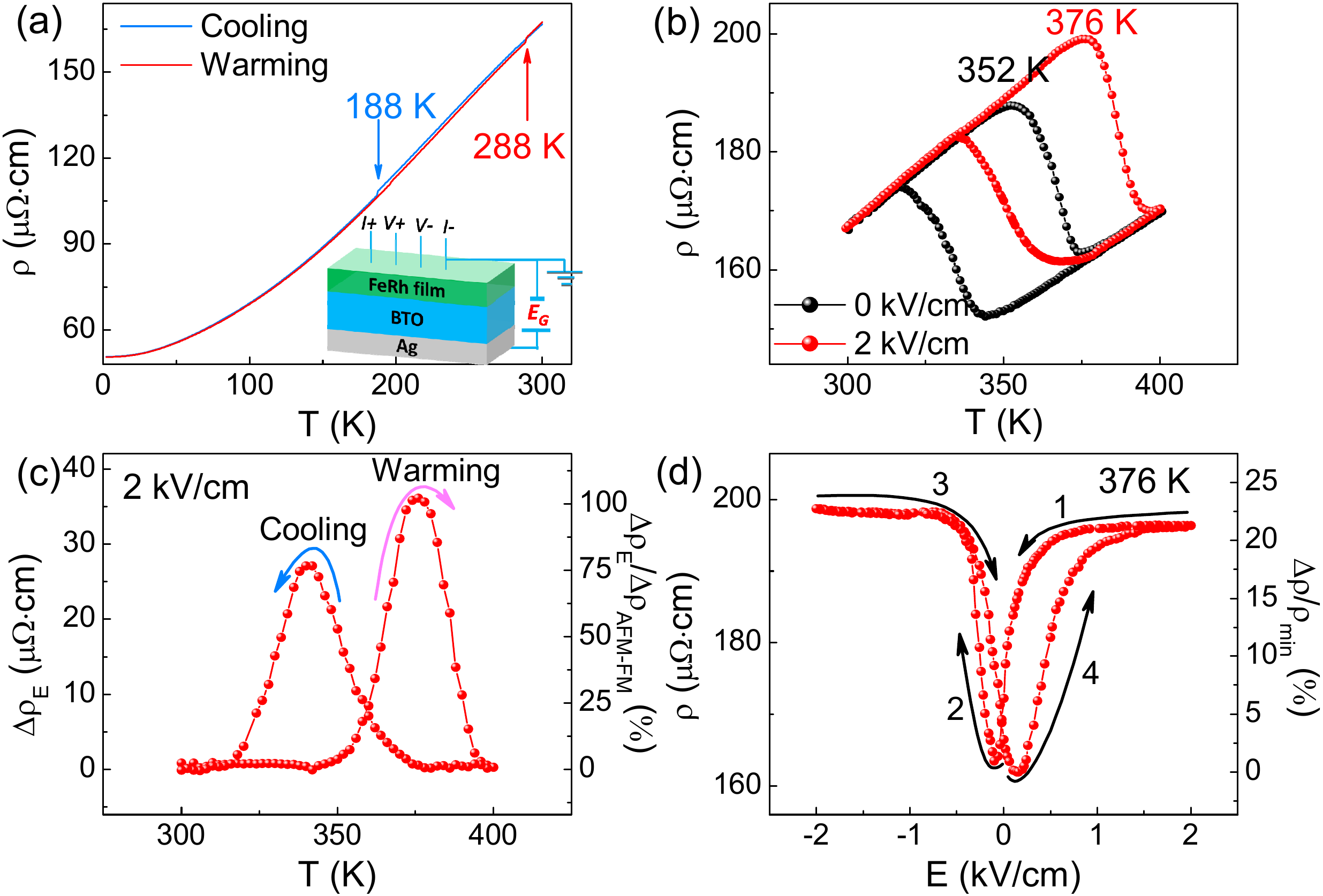}
\caption{\label{fig1} (a) \emph{T}-dependent resistivity ($\rho$-\emph{T}) of an FeRh/BTO heterostructure measured by a d.c. current of 3 mA between 300 and 2 K. Inset: Schematic for the electric gate measurements. (b) $\rho$-\emph{T} curves with and without a gate \emph{E} between 300 and 400 K. (c) Resistivity enhancement $\Delta$$\rho$$_{E}$ under \emph{E} = 2 kV/cm as a function of \emph{T} during warming and cooling. The according ratio of $\Delta$$\rho$$_{E}$ over the resistivity difference between the AFM and FM phases $\Delta$$\rho$$_{AFM-FM}$ $\sim$ 35 $\mu$$\Omega$$\cdot$cm is marked on the right axis.  (d) $\rho$ versus \emph{E} measured at 376 K by scanning \emph{E} from 2 $\rightarrow$ -2 $\rightarrow$ 2 kV/cm. The electroresistance (right axis) is normalized to the resistivity minimum.}
\end{figure}

Electrical transport measurements were conducted in a Quantum Design physical property measurement system combined with Keithley source meters. The resistivity was measured in the linear four-probe geometry with electrical contacts made with Al wires using wire bonding. Figure 1(a) shows \emph{T}-dependent resistivity ($\rho$-\emph{T}) of the FeRh film between 300 and 2 K. The resistivity is $\sim$165 $\mu$$\Omega$$\cdot$cm at 300 K, comparable with previously reported values for thin films [30,39]. The similar signature of the BTO tetragonal-orthorhombic structural transition appears as a subtle kink at $\sim$288 K in the warming-up $\rho$-\emph{T} curve. In addition, the low-\emph{T} orthorhombic-rhombohedral structural transition is observable at $\sim$188 K in the cooling-down $\rho$-\emph{T} curve. However, the effect of BTO structural transitions on the low-\emph{T} FeRh resistivity is overall rather weak, suggesting that the single AFM phase in FeRh is not very sensitive to strain.

Above 300 K, a resistivity drop occurs at $\sim$352 K (defined as the resistivity peak \emph{T}) upon heating in the $\rho$-\emph{T} curve [black in Fig. 1(b)] without a gate \emph{E}, consistent with the onset of the AFM-FM transition in Fig. S1(d). The thermal hysteresis in the $\rho$-\emph{T} curve is similar to in the \emph{M-T} curve as well. With a perpendicular \emph{E} = 2 kV/cm applied to the BTO substrate (silver paint was utilized as the bottom electrode), the onset \emph{T} of the AFM-FM transition shifts to $\sim$376 K (the red curve). The transition \emph{T} enhancement of 24 K is comparable with that of the FeRh/BTO heterostructures in [29]. The out-of-plane \emph{E} switches the in-plane \emph{a}-domains of the BTO substrate to out-of-plane \emph{c}-domains, giving rise to in-plane compressive strain for the FeRh film. It is similar to the FeRh/PMN-PT case [30], where the in-plane compressive strain resulting from domain switching shifts the AFM-FM transition to a higher \emph{T}.

In Fig. 1(b), a significant resistivity modulation by the \emph{E} can only be seen around the PTs during both warming and cooling. While warming up, the resistivity modulation is negligible in the predominant AFM phase below 352 K and the full FM phase above 395 K. This implies the important role of phase instability around the PTs, where the free energies of the AFM phase and the FM phase are quite close to each other [40,41] so that the \emph{E}-induced compressive strain can easily induce the FM-AFM phase transformation. To further analyze the effect of the \emph{E} on the resistivity of FeRh, resistivity enhancement $\Delta$$\rho$$_{E}$ is extracted from the $\rho$-\emph{T} curves in Fig. 1(b) as a function of \emph{T} for warming and cooling. As plotted in Fig. 1(c), $\Delta$$\rho$$_{E}$ peaks at 376 and 340 K during warming and cooling, respectively. The ratio of $\Delta$$\rho$$_{E}$/$\Delta$$\rho$$_{AFM-FM}$, where $\Delta$$\rho$$_{AFM-FM}$ is the resistivity difference between the AFM phase and the FM phase, could serve as an efficacy indicator for the \emph{E}-induced FM-AFM phase transformation. $\Delta$$\rho$$_{AFM-FM}$ can be estimated by deducting the metallic $\rho$-\emph{T} background of the FeRh film in the FM phase from the original $\rho$-\emph{T} curve and is $\sim$35 $\mu$$\Omega$$\cdot$cm in this case (supplemental material). As marked on the right axis of Fig. 1(c), the $\Delta$$\rho$$_{E}$/$\Delta$$\rho$$_{AFM-FM}$ ratio reaches $\sim$100\% at 376 K during warming, which implies a full magnetic phase switch from the FM phase to the AFM phase.

Accordingly, isothermal electroresistance (\emph{ER}) at 376 K was examined. A maximum \emph{ER} ratio of $\sim$22\% is obtained at $\pm$2 kV/cm as shown in Fig. 1(d). The obvious butterfly feature in the \emph{$\rho$-E} curve is in accordance with the resistivity modulation induced by the piezoelectric strain in ferroelectric BTO rather than the electrostatic effect. In addition, an FeRh thin film has a carrier density of $\sim$4.5$\times$10$^{22}$ cm$^{-3}$ in its FM phase [42], similar to usual transition metals. The Thomas-Fermi screening length in the semi-classical model can be expressed as $\lambda_{TF}$ = (2$\varepsilon_{0}$$\varepsilon_{r}$\emph{E}$_{F}$/3\emph{ne}$^{2}$)$^{1/2}$, where $\varepsilon_{0}$ is the vacuum permittivity, $\varepsilon_{r}$ is the dielectric constant, \emph{E}$_{F}$ is the Fermi energy, \emph{n} represents the carrier density and \emph{e} is the electron charge. The Fermi energy can be denoted as \emph{E}$_{F}$ = ($\hbar^{2}$/2\emph{m}$_{e}$)(3$\pi^{2}$\emph{n})$^{2/3}$ with \emph{m}$_{e}$ being the electron mass. Considering a static dielectric constant of 15 for FeRh in its FM phase [43], $\lambda_{TF}$ is estimated to be $\sim$2 {\AA} for 376 K in our case. Compared with the overall thickness of 35 nm, the electrostatic effect on the resistivity modulation is therefore negligible.

\begin{figure}
\includegraphics[width=3.4in]{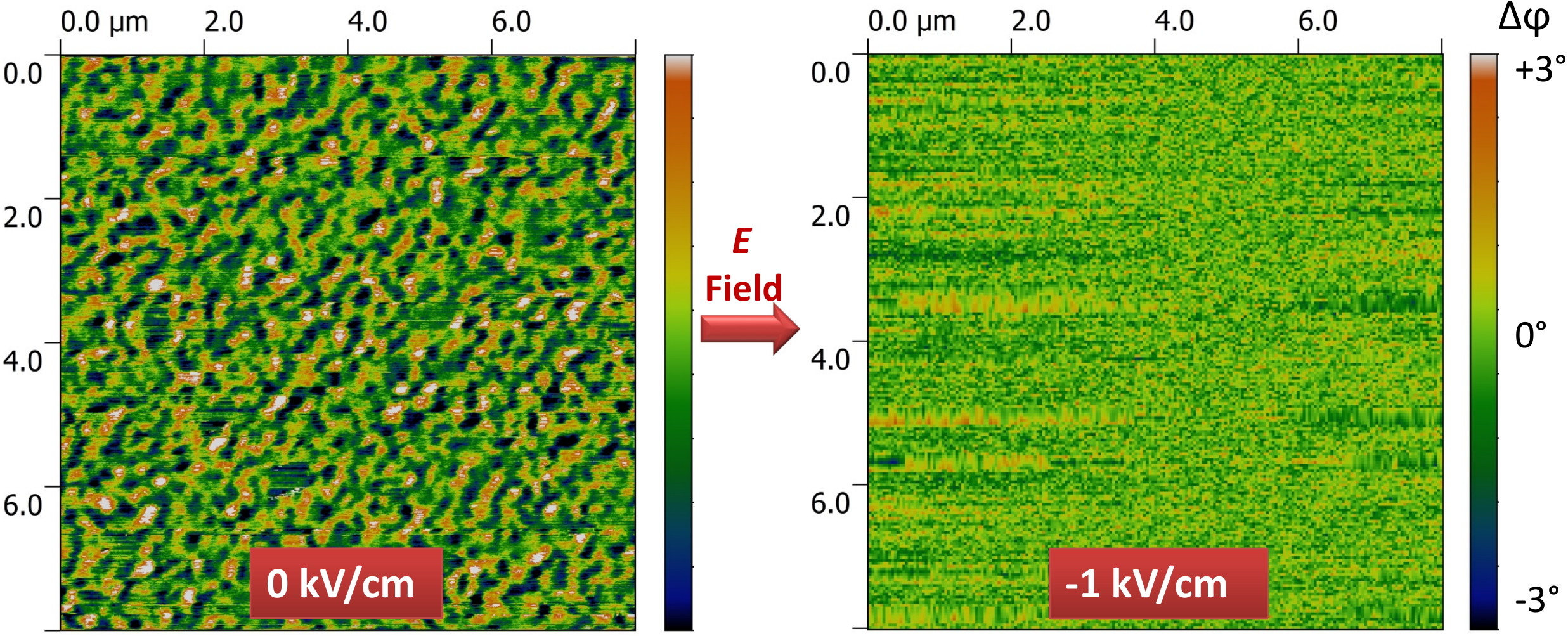}
\caption{\label{fig2} MFM phase images (8$\times$8 $\mu$m$^{2}$) of an FeRh/BTO heterostructure collected at 376 K. Left: MFM image of an FeRh film surface region after heating up to 376 K with zero \emph{E}. Right: MFM image of the same region after an \emph{E} = -1 kV/cm is applied perpendicularly to the BTO substrate at the same \emph{T} $\thicksim$ 376 K. The false color stands for the phase shift $\Delta\varphi$, which reflects the strength and orientation of out-of-plane magnetic moment (orange color represents up moment and blue color down). Green color $\Delta\varphi$ $\thicksim$ 0 corresponds to negligible moment.}
\end{figure}

To directly visualize the effect of an \emph{E} applied on the BTO substrate on magnetic properties of an FeRh thin film, an MFM system with both heating and voltage capabilities was utilized. FeRh exhibits negligible in-plane magnetocrystalline anisotropy [25,26] and the easy axis lies in-plane in thin films. Nevertheless, out-of-plane moment in its FM phase can still be detected by both SQUID and MFM due to the large magnetization in FeRh [30]. After heating an FeRh/BTO heterostructure to 376 K with zero \emph{E}, MFM scans were performed to image magnetic domains of the FeRh film. An MFM phase image collected at 376 K is shown in the left panel of Fig. 2, where the absolute value of the relative phase shift $\mid\Delta\varphi\mid$ is proportional to local magnetic field strength, thus an indicator of local moment. Magnetic domains are uniformly distributed, corresponding to the FM phase in the FeRh film. However, upon an \emph{in-situ} \emph{E} = -1 kV/cm applied to the BTO substrate, the magnetic domains disappear (right panel of Fig. 2). The possibility of moment re-orientation from out-of-plane to in-plane is ruled out by the small anisotropic magnetoresistance effect ($\sim$0.3\% at 376 K, supplemental material), and therefore the \emph{E}-induced domain change together with the large resistivity modulation manifests the full magnetic phase switch. After further heating the sample to 400 K with the \emph{E} = -1 kV/cm, FM domains show up again (supplemental material), demonstrating the \emph{T}-driven AFM-FM transition and also consistent with the negligible resistivity modulation above 395 K in Fig. 1(c). In addition, ferroelectric domains of the BTO single crystals are of tens to hundreds $\mu$m, much larger than the FM domain size in FeRh.

\begin{figure}
\includegraphics[width=3.4in]{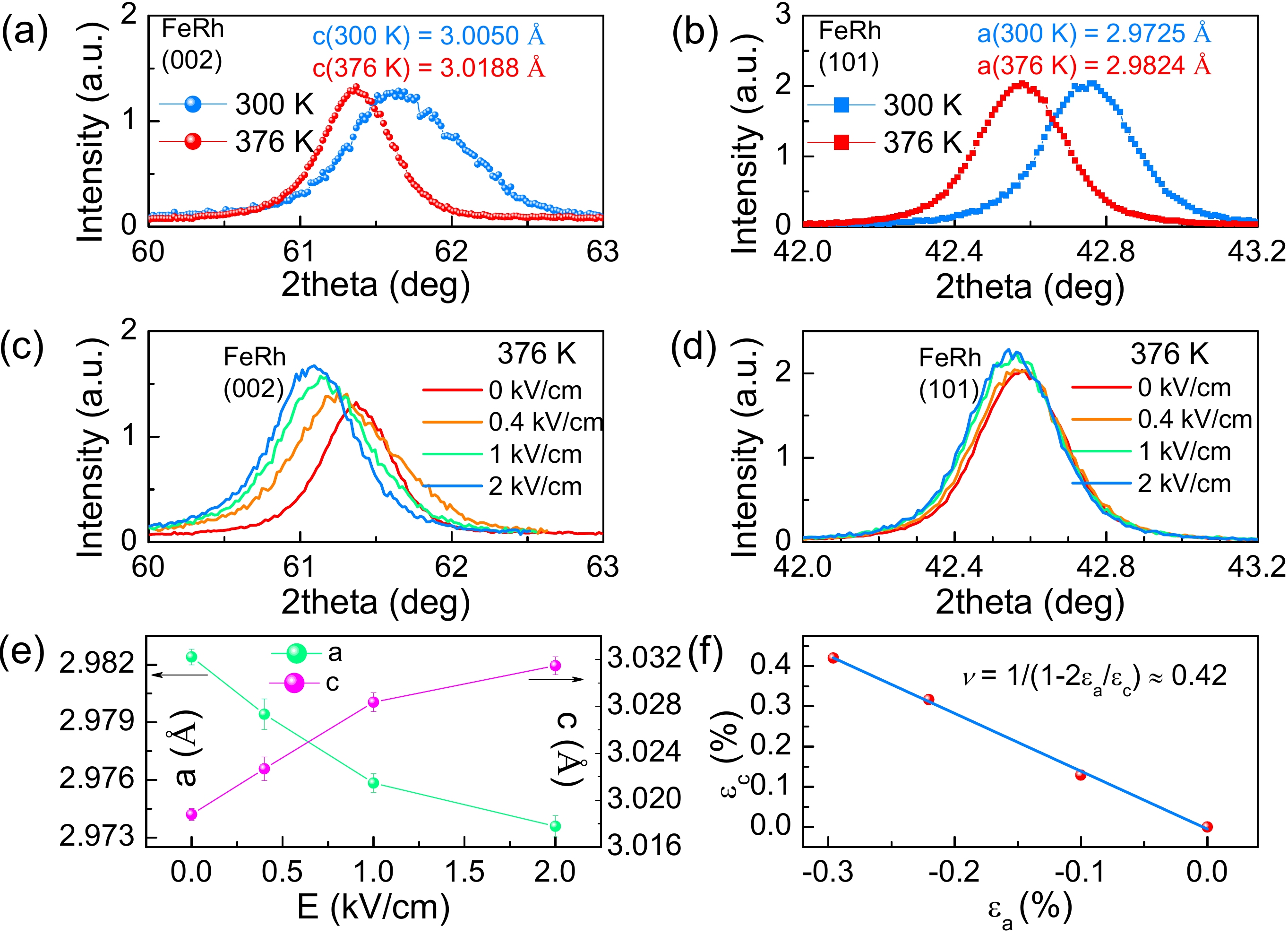}
\caption{\label{fig3} (a) FeRh (002) and (b) (101) peak $\theta$-2$\theta$ scans at 300 and 376 K for an FeRh/BTO heterostructure. (c) FeRh (002) peak and (d) (101) peak $\theta$-2$\theta$ scans under different \emph{E}-fields at 376 K. (e) Lattice constants \emph{a} and \emph{c} of the FeRh film calculated from (c) \& (d). (f) Linear fitting of the out-of-plane strain $\varepsilon_{c}$ versus in-plane strain $\varepsilon_{a}$ for the FeRh film.}
\end{figure}

To quantify the average lattice change of an FeRh/BTO heterostructure during the \emph{E}-induced MPT, x-ray diffraction (XRD) measurements with \emph{in-situ} \emph{E}-fields were carried out at different temperatures. As shown in Figs. 3(a) \& (b), FeRh (002) and (101) peaks both shift to smaller angles upon heating to 376 K, at which the film transits into the FM phase. Compared with \emph{a} = 2.9725 {\AA} and \emph{c} = 3.0050 {\AA} in the dominant AFM phase at 300 K, the FeRh lattice expands significantly in the FM phase, with \emph{a} = 2.9824 {\AA} and \emph{c} = 3.0188 {\AA} at 376 K. Although the tetragonality \emph{c/a} changes only by $\sim$0.12\%, from 1.0110 at 300 K to 1.0122 at 376 K, the resulting FeRh unit cell volume variation is as large as $\sim$1.13\%, from 26.5514 {\AA}$^{3}$ at 300 K to 26.8513 {\AA}$^{3}$ at 376 K. These describe a uniform and large volume expansion across the \emph{T}-driven MPT similar to in bulk [14,15].

During heating, the BTO (002) peak gradually shifts to higher angles and the BTO (020)/(200) peak shifts to low angles with the $\emph{a}$-to-$\emph{c}$ domain ratio varying as well (supplemental material). At 376 K, the $\emph{a}$-to-$\emph{c}$ domain ratio increases to $\sim$56\%/44\% and an out-of-plane \emph{E} = 2 kV/cm applied to the BTO substrate switches almost all the \emph{a}-type domains to \emph{c}-type (supplemental material). During this electrical switching process, the lattice of the FeRh film is modified. Figures 3(c) \& (d) show the FeRh film (002) peak and (101) peak under various \emph{E}-fields, respectively. The corresponding lattice constants are summarized in Fig. 3(e), where \emph{a} decreases while \emph{c} increases with \emph{E} as a consequence of the in-plane compressive strain during BTO domain switching.

Under 2 kV/cm, \emph{a} and \emph{c} change to 2.9736 {\AA} and 3.0315 {\AA} from 2.9824 {\AA} and 3.0188 {\AA} at 0 kV/cm, corresponding to the in-plane and out-of-plane strain $\varepsilon_{a}$ and $\varepsilon_{c}$ to be -0.30\% and 0.42\%, respectively. Accordingly, the tetragonality remarkably increases by $\sim$0.72\%, from 1.0122 at 0 kV/cm to 1.0195 at 2 kV/cm. The \emph{E}-induced FeRh unit cell volume variation is only $\sim$-0.17\%, from 26.8513 {\AA}$^{3}$ at 0 kV/cm to 26.8054 {\AA}$^{3}$ at 2 kV/cm, which is far smaller than the volume difference between the AFM and FM phases distinguished by \emph{T}. These indicate the dominant role of tetragonality modulation in the \emph{E}-induced MPT in epitaxial FeRh thin films. In addition, the linear fitting of $\varepsilon_{c}$/$\varepsilon_{a}$ [Fig. 3(f)] yields the Poisson's ratio $\nu$ $\thicksim$ 0.42 according to $\nu$ = 1/(1-2$\varepsilon_{a}$/$\varepsilon_{c}$) [44], much larger than the bulk value 0.31 [29].

Similar to the CMR effect in manganites [5,6], the AFM phase in FeRh can be converted to the FM phase by an external magnetic field \emph{B} [45] as well, resulting in giant magnetoresistance [46-49]. The critical magnetic field \emph{B}$_{C}$ for driving the AFM-FM transition is strongly \emph{T}-dependent. At 0 K, \emph{B}$_{C}$ is $\sim$30 T [40], the characteristic energy of which can be associated with the free energy difference $\Delta$\emph{F}$_{FM-AFM}$ between the FM and AFM phases at the ground state, \emph{i.e.}, $\Delta$\emph{F}$_{FM-AFM}$(0 K) $\thicksim$ $\hbar\omega$ = $\hbar$\emph{eB}$_{C}$(0 K)/\emph{m}$_{e}$ $\approx$ 3.4 meV with $\omega$ being the cyclotron frequency. \emph{B}$_{C}$ and $\Delta$\emph{F}$_{FM-AFM}$ become smaller at higher \emph{T}s and approach zero around the PT as both experimentally investigated [40,50] and theoretically calculated [41]. That indeed describes the phase instability around the PT in FeRh, where the free energies of the two phases are comparable and therefore the phase transformation can be induced by small external stimuli. Specifically, the slight lattice modulation of FeRh films results in the isothermal MPT in this work.

It is indeed interesting to compare the FeRh/BTO system with the Fe/BTO system in the recent study [51]. Both systems are capable of electric switching of ferromagnetism between full "on" and "off". In the case of FeRh grown on BTO single crystals, epitaxial interfaces mediate piezoelectric strain from BTO single crystals to FeRh films, and such a dynamic strain triggered by external $\emph{E}$-fields is a long-range effect, elastically uniform and reversible even for very thick (\eg, 600 nm) epitaxial films [52,53]. In the Fe/BTO system, the electric switch of magnetism is well confined at the interface, where the polarization (or ion displacement) in BTO films controls the exchange coupling constants of an ultrathin interfacial FeO$_{x}$ layer, leading to magnetic state change during electric polarization switching.

In summary, a giant \emph{ER} effect is achieved in FeRh/BTO heterostructures as a result of the isothermal piezoelectric-strain-induced MPT, which is superior for low-energy applications. Unlike the \emph{T}-driven MPT, the tetragonality modulation in FeRh films plays a dominant role. This work is expected to motivate more experimental and theoretical studies for the fields of both FeRh and multiferroics. In addition, as the study on the CMR systems has been intensively pursued over the last two decades, the similarity between FeRh and manganites in leading to giant physical effects would open new paths for exploring FeRh and other mixed-phase systems exhibiting phase instability.

\begin{acknowledgments}
The work at Berkeley was supported in part by FAME, one of six centers of STARnet, a Semiconductor Research Corporation program sponsored by MARCO and DARPAwas supported by the SRC-FAME program, and Nanosystems Engineering Research Center for Translational Applications of Nanoscale Multiferroic Systems, Cooperative Agreement Award EEC-1160504 (ZQL). The work at ORNL was sponsored by the Laboratory Directed Research and Development (LDRD) Programs of ORNL managed by UT-Battelle, LLC (ZQL and LL); by the US Department of Energy (DOE), Office of Science, Basic Energy Sciences, Materials Sciences and Engineering Division (LL, ASS, TZW, and HNL); and under US DOE grant DE-SC0002136 (ATW). Part of the work was performed at ORNL's Center for Nanophase Materials Sciences, which is DOE Office of Science User Facility. We thank Dr. J. M. Black and Dr. N. Balke in CNMS-ORNL for their help on setting up MFM measurements.
\end{acknowledgments}

\end{document}